# Deep learning methods in speaker recognition: a review

Dávid Sztahó, György Szaszák and András Beke

Department of Telecommunication and Media Informatics, Budapest University of Technology and Economics, Budapest, Hungary

E-mail: sztaho@tmit.bme.hu, szaszak@tmit.bme.hu, beke.andras@gmai.com

**Abstract**

This paper summarizes the applied deep learning practices in the field of speaker recognition, both verification and identification. Speaker recognition has been a widely used field topic of speech technology. Many research works have been carried out and little progress has been achieved in the past 5-6 years. However, as deep learning techniques do advance in most machine learning fields, the former state-of-the-art methods are getting replaced by them in speaker recognition too. It seems that DL becomes the now state-of-the-art solution for both speaker verification and identification. The standard x-vectors, additional to i-vectors, are used as baseline in most of the novel works. The increasing amount of gathered data opens up the territory to DL, where they are the most effective.

## 1 Introduction

Speaker identification (SI) and verification (SV) have a still growing literature, due to their importance in speech technology. It is a popular research topic with various applications, such as security, forensics, biometric authentication, speech recognition and speaker diarization (Hansen and Hasan, 2015). Due to the high number of studies in the field, a great many methods have come up, so state-of-the-art in the field is quite mature, but also versatile, hence hard to overview.

Nowadays, as the popularity of deep learning (DL) is constantly rising due to easy accessible software and affordable hardware solutions, it began to infiltrate every topic, where machine learning is applicable. So, it is only natural, that experts and scientists began to use deep learning is speaker recognition (SR). The aim of this study is to review the deep learning methods that are applied in speaker identification and verification tasks from the earliest to the latest solutions.

First, it is necessary to clarify the definition of speaker identification and verification, since these tasks are generally referred to when performing speaker recognition (Hansen and Hasan 2015). Speaker identification is the task to identify an unknown speaker from a set of already known speakers: find the speaker who sounds closest to the test sample. When all speakers within a given set are known, it is called closed-set (or in-set) scenario. Alternatively, if the set of known speakers may not contain the potential test subject, it is called open-set (or out-of-set) speaker identification.

In speaker verification, the task is to verify if a speaker, who claims to be of an identity, really is of the identity. In other words, we have to verify if the subject is really who he or she says to be. This means comparing two speech samples/utterances and deciding if they are spoken by the same speakers. This is - in general speaker verification practice - usually done by comparing the test sample to a sample of the given speaker and a universal background model (Reynolds et al., 1995).

Both speaker identification and verification have their use-cases and in most cases, similar or same methods are required. Therefore, in this review, we examine the DNN methods for both scenarios, indicating the given task in every mentioned literature every time.

We focus only on DL methods in the field, as current state-of-the-art builds almost exclusively on top of neural architectures. For a speaker recognition tutorial, we recommend the work of (Hansen and Hasan 2015). Due to the extensive nature of the field of deep learning, it is beyond the scope of this paper to give a detailed introduction about it. The methods are discussed with the assumption that the reader has usable knowledge in the field.

**2. Databases**

Like in many speech technology (and other machine learning) related topics, the question of the used database is crucial. Developed methods can be evaluated and compared only if the same testing circumstances (from a machine learning point of view) are used. It is hard to say that an approach performs better, if it is evaluated on a different set or corpus. Therefore, the selection of the training and evaluation datasets require taking different considerations into account. There are numerous databases that are created and used in the field of speaker recognition, identification and verification. In Table I, presently available corpuses are listed along with their different properties that are found publicly. There are some datasets that are free, some are freely available for research only.

Corpuses that are created mainly for automatic speech recognition (ASR) can also be used to train (and evaluate) SR methods, however, most researches use datasets that focus especially to the field of speaker identification and verification. The main difference is the number of speakers contained in the database. Databases made for speech recognition typically contain less speakers. Speech recognition needs much more speech data in order to train phoneme models, but it often comes with lower speaker number. In contrast, speaker recognition needs as many speakers as possible, with less needed recorded material from each speaker. Also, recruiting many speakers is a more challenging job that requires more effort and is time consuming. The most often used corpuses for speech recognition (such as TIMIT (Garofolo et al., 1993), WSJ (Marcus et. al, 1993), RSR2015 (Larcher et al., 2012), CHiME 2013 (Vincent et al., 2013), VCTK (Veaux et al., 2017)) have a few hundred speakers, whereas Librispeech (Panayotov et. al, 2015), VoxCeleb (Nagrani et. al, 2017), NIST SRE (Greenberg et. al, 2017) datasets contain thousands of speakers. Of course, these large corpuses likely contain audio samples with various background noises, signal-to-noise ratios, recording setups and equipment. Therefore, they are suitable for machine learning aspects, but may not from a linguistic point of view, in which case a more homogenous and a clean recording quality is necessary.

The largest corpus especially made for SR tasks is VoxCeleb2 (Chung et. al, 2018). It is a recent extension to its previous version (VoxCeleb). It contains samples from more than 6000 speakers that are downloaded from Youtube. Thus, its sound quality varies largely. In contrast, LibriSpeech is a clean, good quality corpus. It is created from audiobooks, therefore maybe less suitable from real-world usage point of view, but appropriate for evaluating SR methods and features. NIST SRE datasets are also a huge collection of speaker samples, but recorded through telephone line quality. Thus, suitable for evaluations in yet another usage environment. A noisy and band limited quality makes the SR task harder, therefore is more suitable to make a comparison between SR methods.

Although most corpuses contain English speech material, there are also other languages available. Some even contain multilingual content (see Table I). Another important aspect is if the given corpus has any additional segmentation or transcription included. If so, a more subtle analysis can be carried out (for example, using only given phonemes or partitioning the corpus into chunks with different sizes).

Table I shows all the corpuses that are used in the literature that is reviewed throughout the present paper.

**3. Short History: GMM-UBM and i-vector**

*3.1. GMM-UBM*

The first automatic speaker identification method was based on Gaussian mixture models (GMM) (Hansen and Hasan, 2015; Reynolds et al., 1995). GMM is a combination of Gaussian probability density functions (PDFs) that are commonly used to model multivariate data. It does not only cluster data in an unsupervised way, but also gives its PDF. Applying GMM to speaker modelling provides the speaker specific PDF, from which probability score can be obtained. Thus, testing a sample with an unknown label, based on the probability scores of the speaker GMMs, a decision can be made.

A GMM is a mixture of Gaussian PDFs parameterized by a number of mean vectors, covariance matrices and weights:

$$f(x_n|\lambda) = \sum_{g=1}^{M} \pi_g N(x_n|\mu_g, \Sigma_g)$$

where $\pi_g, \mu_g$ and $\Sigma_g$ indicate the weight, mean vector and covariance matrix of the $g^{th}$ mixture component. For a sequence of acoustic features ($X = \{x_n | n \in 1 \dots T\}$), the probability of observing these features is computed as

$$p(X|\lambda) = \prod_{n=1}^{T} p(x_n|\lambda).$$

For speaker verification scheme, a slightly different approach was developed (Hansen and Hasan, 2015). Beside the claimed speaker's model, an alternate model is necessary, which represents an 'opposing' model. This alternate model is called the universal background model (GMM-UBM). The GMM-UBM represents all others than the target speaker and it is trained on a large number of speaker samples. It was first used in (Reynolds et al., 2000). Later, UBM was used as an initial model to the speaker models: rather than training GMMs on speaker data directly, the specific speaker models were created by adapting a prior UBM (Gauvain and Lee 1994). In the GMM-UBM scheme, H0and H1in (1) (see later in Section about LR Test) are represented by speaker dependent GMM and the GMM-UBM, respectively.

### 3.2. GMM Supervectors

Because speech samples could have different durations, much effort was put into developing methods that can obtain a fixed number of features from samples with variable lengths. One of the methods that performed the best in speaker recognition is forming GMM supervectors (Campbell et. al, 2006). Supervectors are created by concatenating the parameters of the GMM (the mean vectors). This fixed length 'supervector' is than fed to a obligable machine learning technique. Before deep neural networks began to take much attention, support vector machines (SVM) (Cortes and Vapnik, 1995) were found to be the best performing technique.

### 3.3. i-vector

Also, before the deep neural networks era, the state-of-the-art speaker recognition method was the i-vector approach (Dehak et. al, 2009a; Dehak et. al, 2009b; Dehak et. al, 2011). In this model, factor analysis (FA) was used to compute a speaker- and session-dependent GMM supervector:

$$m_{s,h} = m_0 + Tw_{s,h},$$

where $m_0$ is the GMM-UBM supervector, $T$ is the speaker and channel factor, called total variability space and $w_{s,h} \sim N(0,1)$ are hidden variables, called total factors. The total factors are not observable, but can be estimated using FA. These total factors than can be used as features to a classifier, and came to be known as *i-vector*s (short for identity vector). The i-vector approach can be considered as a dimensionality reduction technique of the GMM supervector.

### 3.4. LR Test

In speaker verification, the decision if a test sample belongs to a certain speaker is generally given by the likelihood-ratio test (LR test) (Hansen and Hasan, 2015). There are two hypotheses for an observation O:

$$H_0: O \text{ is from speaker } s$$
$$H_1: O \text{ is not from speaker } s$$

In most of the approaches these cases are represented by a certain model parameterized by sand 1, respectively. For a given set of observations $X = \{x_n | n \in 1 \ldots T\}$, the LR test is applied by evaluating the following ratio:

In most of the approaches these cases are represented by a certain model parameterized by $\lambda_s$ and $\lambda_1$, respectively. For a given set of observations $X = \{x_n | n \in 1 \ldots T\}$, the LR test is applied by evaluating the following ratio:

$$p(X|\lambda_s) \geq \tau \text{ accept } H_0$$
$$p(X|\lambda_1) > \tau \text{ reject } H_0,$$

where $\tau$ is the threshold of the decision. Commonly, the LR Test is computed by using logarithmic probabilities (log-LR):

$$\Lambda(X) = \log p(X|\lambda_s) - \log p(X|\lambda_1). \quad (1)$$

## 4 Speaker verification measurements

In speaker recognition (especially in verification) there are two kinds of similarity measures that are commonly used to compute the probabilities if a test observation is from the target speaker or not. Almost all novel DL approaches use these measures (in speaker verification schemes): cosine distance of vectors and PLDA (probabilistic linear discriminant analysis).

### 4.1. Cosine Distance

The cosine distance is simply computing the normalized dot product of target and test i-vectors ($w_{target}$ and $w_{test}$), which provides a match score:

$$CDS(w_{target}, w_{test}) = \frac{w_{target} \cdot w_{test}}{||w_{target}|| \cdot ||w_{test}||}$$

### 4.2. PLDA

LDA (linear discriminant analysis) (Bishop, 2006) is used to find orthogonal axes for minimizing within-class variation and maximizing between-class variation. PLDA, as an extension of LDA (Tipping and Bishop, 1997; Ioffe, 2006), is a probabilistic approach to the same method.

Generally, PLDA was applied to compare i-vectors. Of course, PLDA is capable to be applied to any vectors. Therefore, it can be used in new DL approaches, where i-vectors are replaced with their deep learning alternatives. Here, we give a brief description using the traditional i-vector approach.

Given a set of $d$ dimensional length-normalized i-vectors $X = \{x_{ij}; i = 1, \ldots, N; j = 1, \ldots, H_i\}$ obtained from N training speakers (each has $H_i$ i-vectors), i-vectors can be written in the following form:

$$x_{ij} = \mu + Wz_i + \epsilon_{ij}$$
$$x_{ij}, \mu \in R^D, W \in R^{DxM}, z_i \in R^M, \epsilon_{ij} \in R^D,$$

where $Z = \{z_i; i = 1, \ldots, N\}$ are latent variables, $\omega = \{\mu, W, \Sigma\}$ are model parameters, $W$ is a $DxM$ matrix (called factor loading matrix), $\mu$ is the global mean of $X$, $z_i$'s are called the speaker factors and $\epsilon_{ij}$'s are Gaussian distributed noise with zero mean and $\Sigma$ covariance.

Given a test i-vector $x_t$ and a target-speaker i-vector $x_s$, the LR score can be computed:

$$S_{LR}\{x_t, x_s\} = \frac{P(x_s, x_t | same\ speaker)}{P(x_s, x_t | different\ speaker)}$$

$$S_{LR}\{x_t, x_s\} = \frac{\int p(x_s, x_t, z | \omega) dz}{\int p(x_s, z | \omega) dz_s \int p(x_t, z | \omega) dz_t}$$

$$S_{LR}\{x_t, x_s\} = \frac{\int p(x_s, x_t | z, \omega) p(z) dz}{\int p(x_s | z_s, \omega) p(z_s) dz_s \int p(x_t | z_t, \omega) p(z_t) dz_t}$$

$$S_{LR}\{x_t, x_s\} = \frac{N([x_s^T\ x_t^T]|[\mu^T\ \mu^T], \widehat{W}\widehat{W}^T + \widehat{\Sigma})}{N([x_s^T\ x_t^T]|[\mu^T\ \mu^T], diag\{WW^T + \Sigma, WW^T + \Sigma\})} \quad (2)$$

where $\widehat{W} = [W^T\ W^T]^T$ and $\widehat{\Sigma} = diag\{\Sigma, \Sigma\}$. Using Eq. (2) and the standard formula for the inverse of block matrices [Petersen and Pedersen, 2008], the log-likelihood RL score is given by (Ioffe, 2006):

$$S_{LR}(x_s, x_t) = const + x_s^T Q x_s + x_t^T Q x_t + 2 x_s^T P x_t$$

where

$$P = \Lambda^{-1} \Gamma (\Lambda - \Gamma \Lambda^{-1} \Gamma)^{-1}; \Lambda = WW^T + \Sigma$$
$$Q = \Lambda^{-1} - (\Lambda - \Gamma \Lambda^{-1} \Gamma)^{-1}; \Gamma = WW^T.$$

## 5 Deep Learning in Speaker Recognition

Generally, deep learning in speaker recognition has two major directions. One approach is to replace the i-vector calculation mechanism with a deep learning method as feature extraction. These works train a network on speaker samples using acoustic features (such as MFCCs or spectra) as inputs and speaker IDs as target variable and commonly use the output of an internal hidden layer as i-vector alternative and apply cosine distance or PLDA as decision making. The other main strategy is to use deep learning for classification and decision making, like replacing the cosine distance and PLDA with a discriminating deep network.

The performance of automatic speaker recognition systems are commonly evaluated by equal error rate (EER) and decision cost function (DCF). Equal error rate (EER) is a biometric security system algorithm used to predetermines the threshold values for its false acceptance rate and its false rejection rate (Van Leeuwen and Brummer, 2007; Hansen and Hasan, 2015). When the rates are equal, the common value is referred to as the equal error rate. The value indicates that the proportion of false acceptances is equal to the proportion of false rejections. The lower the equal error rate value, the higher the accuracy of the biometric system. Alternatively, the decision cost function takes the prior probabilities of the target speaker occurrences, the proportion of target and non-target speakers into consideration. The detection cost function is a simultaneous measure of discrimination and calibration. Often, the minimum value of the DCF curve is called minDCF.

### 5.1. Deep learning for feature extraction

(Chen and Salman, 2011) is a relatively early work in deep feature extraction, in which bottleneck features (speaker models) are created using a deep neural network with multiple subsets. Each subset is a deep autoencoder originally proposed in (Hinton and Salakhutdinov, 2006). A hybrid learning strategy is proposed: the weights of the middle layer are shared across multiple inputs (adjacent frames) by a cost function:

$$L(x_1, x_2, \theta) = [L_R(x_1; \theta) + L_R(x_2; \theta)] + L_E(x_1, x_2; \theta)$$

where $L_R(x_i; \theta)$ is the loss of the network for input $i$, and $L_E(x_1, x_2; \theta)$ is a loss function optimized for learning the same speaker representation (model) at the layer, from which the speaker model features are extracted. For the experiments TIMIT, NTIMIT, KING, NKING, CHN and RUS dataset are used. According to the results, the proposed method outperformed the GMM-UBM baseline system in the case of all datasets.

#### 5.1.1. d-vector

There are numerous works that are aimed at extracting hidden layers of a DNN as features (substituting i-vectors). In (Variani et. al, 2014) averaged activations of the last hidden layer of a network with multiple fully connected layers are selected as features, called as 'd-vector' (Fig. 1.). These vectors are later used in the same manner as i-vectors. Speaker verification is done by cosine distance comparison. First, the network is trained by supervised manner, using 13-dimensional perceptual linear predictive (PLP) features with Δ and ΔΔ values appended as frame-level feature vectors. After the training, the output layer is removed and the activations from the last hidden layers are used as features. The experiments were performed on a small footprint text-dependent corpus: 646 speakers speaking the same phrase: 'ok google', multiple times. It was found that the general i-vector system mainly outperforms the newly proposed d-vector. The EERs (score normalized with t-norm) of the best performing setups were 1.21% and 2.00% for i-vector and d-vector, respectively.

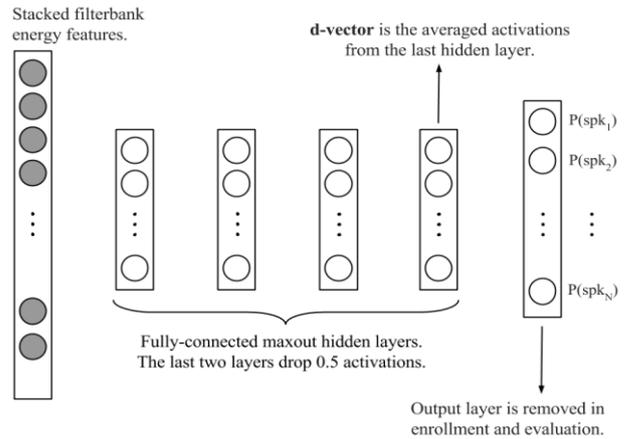

Fig. 1. DNN model in (Variani et al., 2014).

#### 5.1.2. j-vector

The d-vector method was extended in (Chen et. al, 2014) by a multi-task learning approach. The authors state that the intuition is that directly recognizing speaker seems to be hard but in reality different speakers have their own style on each syllable or word. Therefore, using not only the speaker ids, but texts also as targets in a multi-learning setup, may increase the speaker verification performance. The used network is shown in Fig. 2. The applied cost function is the sum of the original loss functions:

$$C([y_1, y_2], [y_1', y_2']) = C_1(y_1, y_1') + C_2(y_2, y_2')$$

where $C_1$ and $C_1$ are two cross-entropy criteria for speakers and texts, $y_1, y_2$ indicate the true labels for speakers and texts individually and $y_1', y_2'$ are the outputs of the two targets. As in the case of the original d-vector, after the supervised training phase, the output layer is removed and the output of the last hidden layer is used as a feature vector, defined as j-vector (joint vector). The experiments were done on the RSR2015 database (Larcher et. al, 2012). The results show that the j-vector outperformed the d-vector approach. The EERs are 21.05% and 9.85% for *d-* and *j-vector*, respectively.

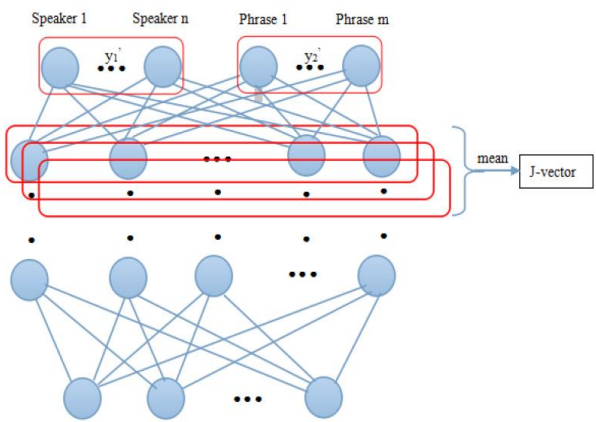

*Fig. 2. Multi-task DNN in (Chen et. al, 2014).*

### 5.1.3. x-vector

Another hidden layer extracted feature vector is called x-vector (Snyder et al., 2018; Fang, 2019). It is based on DNN embeddings, which employs a multiple layered DNN architecture (with fully connected layers) with different temporal context at each layer (which they call 'frames'). Due to the wider temporal context, the architecture is called time-delay NN (TDNN). The TDNN embedding architecture can be seen in Fig. 3. The first five layers operate on speech frames, with small temporal context centered at the current framt *t*. For example, frame3 sees a total of 15 frame, due to the temporal context of the earlier layers. After training to speaker ids as target vectors, the output of layer segment6 ('x-vector') is used as input to a PLDA classifier. The input acoustic features are 24 dimensional filterbanks with 25ms frame size, mean-normalized over a sliding window of up to 3 seconds. The used databases for evaluation are SWBD, NIST SRE 2016 and VoxCeleb. Data augmentation (increasing the amount of samples by adding babble noise, background music and reverb) was applied to various experimental setups. The main results show that x-vector outperforms the general i-vector based system (EERs are 9.23% and 8.00% for i-vector and x-vector, respectively). Using data augmentation, the difference is larger (EERs are 8.95% and 5.86% for i-vector and x-vector, respectively). (Jiang et al., 2019) extends the x-vector framework by so called dilated dense blocks, gate block and transition blocks. These blocks use convolutional layers to cover local features of different spans. On VoxCeleb, the extension results in 0.86% EER decrease in absolute value (from 3.17% to 2.31%). Speaker representations can also be used to change the identity of the speaker. In (Fang, 2019) x-vectors are used for speaker anonymization. The extracted vector values are modified in order to change the speaker characterization and the speech is then re-synthetized, generating anonymized speech.

For short speech utterances, (Kanagasundaram et al., 2019) changed the dimension of the sixth and seventh layer ('segment6' and 'segment7') to 150 in order to adapt to the shorter duration. It was found that the lower dimension of segment 6 and 7 helped in speaker verification in the case of 5 second long utterances, but achieved higher EER on the original long utterances (NIST SRE 2010 dataset was used). On the other hand, (Garcia-Romero et al., 2019) tried to optimize the x-vector system for long utterances (with 2-4 seconds duration) by a DNN refinement approach that updates a subset of the DNN parameters with full recordings and modifies the DNN architecture to produce embeddings optimized for cosine distance scoring. The results show that the method produces lower minDCF (minimum decision cost function), but slightly higher EER than the baseline x-vector approach.

The x-vector was also applied in a multi-task learning scenario (You et al., 2019). Beside the primary task (learning speaker identities), a second task was introduced: learning higher-order statistics of the input vector. By doing so, the system achieved slightly lower EER than the standard x-vector on the NIST SRE16 dataset: 7.79% and 8.03% for multi-task and baseline, respectively.

x-vectors are, in general, incapable of leveraging unlabelled utterances, due to the classification loss over training speakers. (Stafylakis et al., 2019) offers an alternative strategy based on x-vectors to train speaker embedding extractors via reconstructing the frames of a target speech segment, given the inferred embedding of another speech segment of the same utterance. They use a decoder network, to which the embedding vector is attached and by which the network serves as an autoencoder. The proposed decoder loss combined with the standard x-vector architecture and loss (i.e., crossentropy over training speakers) yielded improvement both on SITW and VoxCeleb datasets: ~0.4% improvement in absolute EER compared to the standard x-vector system.

| Layer | Layer context | Total context | Input x output |
|---|---|---|---|
| frame1 | $[t-2, t+2]$ | 5 | 120x512 |
| frame2 | $\{t-2, t, t+2\}$ | 9 | 1536x512 |
| frame3 | $\{t-3, t, t+3\}$ | 15 | 1536x512 |
| frame4 | $\{t\}$ | 15 | 512x512 |
| frame5 | $\{t\}$ | 15 | 512x1500 |
| stats pooling | $[0, T)$ | $T$ | $1500T$x3000 |
| segment6 | $\{0\}$ | $T$ | 3000x512 |
| segment7 | $\{0\}$ | $T$ | 512x512 |
| softmax | $\{0\}$ | $T$ | 512x$N$ |

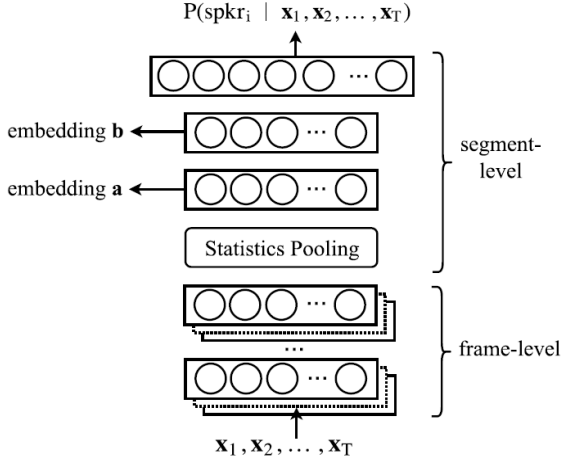

*Fig. 3. x-vector DNN embedding architecture in (Snyder et al., 2018).*

### 5.1.4. End-to-end systems

In order to do speaker verification, the embeddings are extracted and used in a standard backend, e.g., PLDA. Ideally the NNs should however be trained directly for the speaker verification task (Heigold et al., 2016; Rohdin et al., 2018; Gao et al., 2019; (Yun et al., 2019)

Instead of using cosine distance or PLDA classification, (Heigold et al., 2016) applies an end-to-end solution for speaker verification with deep networks to obtain speaker representation vectors, estimation of a speaker model based on up to N enrollment utterances and also for verification (cosine similarity/logistic regression). The architecture is shown in Fig. 4. Both DNN (the same as the network used in d-vector extraction) and LSTMs are applied for speaker representation computation, The network is optimized using the end-to-end loss:

$$l_{e2e} = -log\, p(target)$$

with the binary variable $target \in \{accept, reject\}$, $p(accept) = (1 + e^{-wS(X,spk)-b})^{-1}$ and $p(reject) = 1 - p(accept)$. The value $-b/w$ corresponds with the verification threshold. $S(X, spk)$ is the cosine similarity between the speaker representation and the speaker model. The methods were tested on the 'ok, google' dataset with more than 73M utterances and 80 000 speakers. The results show that the end-to-end architecture performs similar to the d-vector approach if the same feature extractor (DNN) is used. However, LSTM lowered the EERs compared to the DNN solution: EERs are 2.04% and 1.36% for DNN and LSTM, respectively.

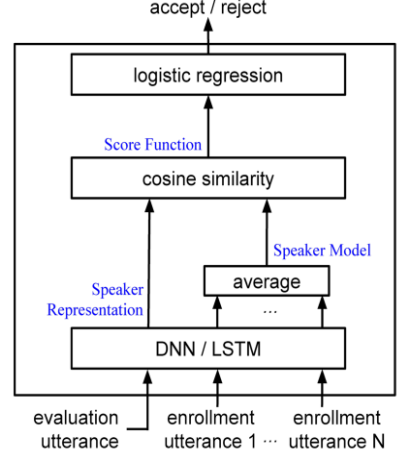

*Fig. 4. End-to-end architecture used in (Heigold et al., 2016).*

Another end-to-end system is proposed in (Yun et al., 2019), where the training was done by triplet loss aided by cosine similarity. A speaker embedding network is fed with raw speech waveform, which produces embedding vectors. This network is pre-trained with LibriSpeech by 1.5-2.0 sec utterance chunks. Then the CHiME 2013 database (Vincent et al., 2013) was used for speaker verification evaluation using specific 2 to 4 keywords only. The keywords were determined by an ASR, which was used in the training of the speaker embedding system in an adversarial way, forcing the embedding vectors to be speaker independent. The results are mixed. The triplet loss and ASR adversarial training did not improve the EER in the 2 keywords case, just when 3 or 4 keywords were examined.

### 5.1.5. Deep belief networks

Deep belief networks (DBN) are another type of deep learning networks that are used in speaker recognition (Ali et al., 2018; Banerjee et al., 2018). Deep belief networks are generative models with numerous layers of latent variables, which are typically binary. Neurons in the same layers are not connected and connection between adjacent layers are undirected. Training of DBNs are hard due to the intractability of inferring the posterior distribution from the hidden (latent) layers. Stacked Restricted Boltzmann Machines (RBMs) can be applied as a DBN architecture (Fig. 5.). For greater details, see (Hinton et al., 2006). The objective of DBN is to learn abstract hierarchical representations of unlabelled input data. In (Banerjee et al., 2018), spectrograms (25 ms window size, 10 ms timestep) have been fed as input speech data after applying PCA transformation to reduce dimensionality. Activations of first and second layers of the RBM were used as features (both separately and together) appended to common MFCC features. After feature extraction, GMM-UBM was used to perform speaker recognition. The authors used the ELSDSR dataset with 22 speakers. Based on the results, the features extracted from the RBM helped the recognition: 90% and 95% final accuracies were obtained by using separate MFCC and mixed MFCC+RBM features, respectively.

(Ali et al., 2018) also uses the same acoustic feature extraction method, but it adds a Bag of Words method in order to convert the data with different lengths into vectors of the same dimensionality (using a k-means clustering technique). SVM was applied as classifier. The experiments were done on the Urdu dataset (Appen, 2007) with ten speakers. Here, also hybrid (MFCC+DBN) features performed the best: 88.6% and 92.60% accuracies were obtained for MFCC and MFCC+DBN features, respectively.

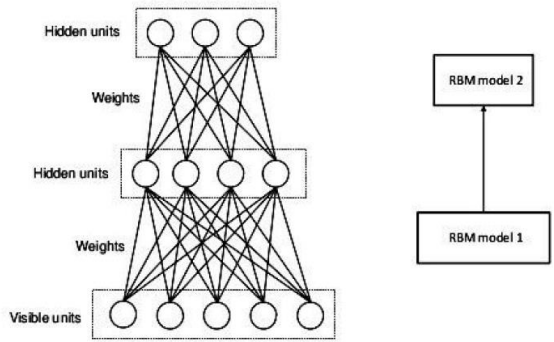

Fig. 5. Structure of the DBN used for extraction of short term spectral features, with two hidden layers, can be visualised as a stack of 2 RBMs (Banerjee et al., 2018)

In (Liu et al., 2015), a widespread evaluation of multiple DNN methods for deep feature extraction are given using deep restricted Boltzmann machines (RBMs), speech-discriminant deep neural network, speaker-discriminant neural network and multi-task joint-learned deep neural networks. RBMs are used in the same way as in the previous section (Ali et al., 2018; Banerjee et al., 2018). A speech discriminant DNN was applied with text labels as training data and triphone states as target. This scenario can be useful in a text-dependent speaker verification task. The outputs of the last hidden layer are used as features. In the case of speaker discriminant DNN, the outputs of the speech discriminant network are changed to speaker IDs. This way, a more speaker specific feature set can be obtained and it is a more natural choice for speaker verification. In the multi-task setup, both previously mentioned (speaker IDs and triphones) outputs are used as targets. A standard i-vector system trained with PLP features was used as baseline (GMM-UBM with cosine similarity). The newly proposed deep features were tested separately and by combining them in various ways on the RSR2015 dataset (Larcher et al., 2012). Compared to the baseline result (1.5% EER), the speaker discriminant and multi-task DNNs achieved the best performances (1.06% and 0.80% EER respectively). The best combination of deep features (concatenating RBM and multi-task features) gave 0.73% EER. Also, with PLDA performed after deep feature extraction gave 0.20% EER for speaker discriminant features.

### 5.1.6. CLNets

In (Wen et al., 2018) a deep corrective learning network (CLNet) is proposed to analyse the independent samples by a recurrent formalism. Each new instance makes a corrective prediction to update the predictions made from prior data. This means that instead of averaging the results for segments of a speaker, an incremental strategy is used. CLNets are applied using convolution layers for speaker verification. NIST SRE 2004-2010 corpora are used for the experiments. By using cosine similarity, ~2.5% lower EER was obtained compared to the standard i-vector system (7.3%, 5.18% and 4.87% EERs for i-vector, standard CNN and CLNets, respectively). However, using PLDA, i-vector performed better.

### 5.1.7. Text dependency

Still, i-vector systems outperform the DNN ones in a text independent scenario (Snyder et al., 2016). So, taking the standard i-vector PLDA system as basis, (Rohdin et al., 2018) proposed an end-to-end DNN that learns sufficient statistics of GMM-UBM and provides i-vectors. In the first part of the network, GMM posteriots are learned by a multiple layered architecture, than the standard i-vectors are used as targets with cosine distance as loss function.

### 5.2. Deep learning for classification

Rather then applying deep feature extraction to exchange the common i-vectors for a more robust and better performing speaker representation, DNNs can also be used to replace the backend systems for scoring and comparison (like PLDA and cosine distance). Such works are more sparse in literature than those related to feature extraction.

### 5.2.1. Variational autoencoder

Variational autoencoder (VAE) (Kingma and Welling, 2013; Rezende et al., 2014) is a generative model for signal (and speech) modelling. It is used in voice conversion (Hsu et al., 2017a; Hsu et al., 2017b), speech recognition and also for speaker recognition (Villalba et al., 2017; Pekhovsky and Korenevsky, 2017). Instead of using just deterministic layers, a VAE consist of stochastic neurons also. The LLR scoring is made by:

$$LR(x_1, x_2) = \frac{P(x_1, x_2|H_{tar})}{P(x_1, x_2|H_{imp})} = \frac{P(x_1, x_2|\theta)}{P(x_1|\theta)P(x_2|\theta)}$$

where $H_{tar}, H_{imp}$ are the hypothesis about the facts that $x_1, x_2$ are related to the same or different speakers respectively and $\theta$ is the parameters of the speaker model. The results showed that VAEs don't seem to be superior to PLDA scoring.

## 5.2.2. Multi-domain features

Using text dependent data to help learning speaker IDs were also used for classification in a speaker recognition task, (Tang et al., 2016) used the output of an ASR to improve the performance of speaker recognition. Fig. 6 shows the proposed multi-task learning. The output of the ASR (phone-posteriors) is fed into the SRE system, and vice versa. The input of each tasks are the extracted frame-level spectra (filterbanks and MFCCs for ASR and SRE, respectively). The experiments were done on the WSJ dataset. Based on the results, the proposed method achieved equal or slightly better EERs, than the i-vector baseline (0.57% and 0.55% for i-vector and multi-task method, respectively).

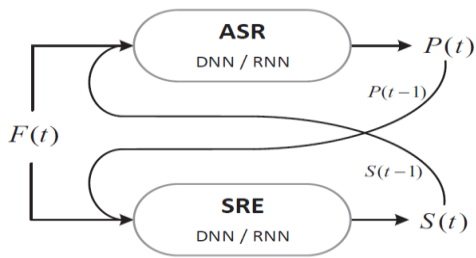

Fig. 6. Multi-task recurrent learning in (Tang et al., 2016) for ASR and SRE. $F(t)$ denotes primary features (e.g., Fbanks), $P(t)$ denotes phone identities (e.g., phone posteriors, high-level representations for phones), $S(t)$ denotes speaker identities (e.g., speaker posteriors, high-level representations for speakers).

## 5.2.3. Replacing UBM with DNN

DNNs can be used to replace the UBM also. Universal deep belief networks (UDBN) (Ghahabi and Hernando, 2017) are used as backend, in which a two-class hybrid DBN-DNN is trained for each target speaker ti increase the discrimination between target i-vector/s and the i-vectors of tóother soeakers (non-targets/impostors). Fig. 7 shows the train/test phases of the proposed method. First, an unsupervised universal DBN is trained, which is then adapted to the target speakers by a special balanced training process. In the test phase, an unknown i-vector is matched to the adapted target i-vectors. Based on evaluation done on NIST SRE 2006 and 2014 datasets, the proposed algorithm did not achieve better performance than the i-vector PLDA baseline method. However, fusing the DNN approach with the PLDA (i-vector) method, revealed better performance than the i-vector alone.

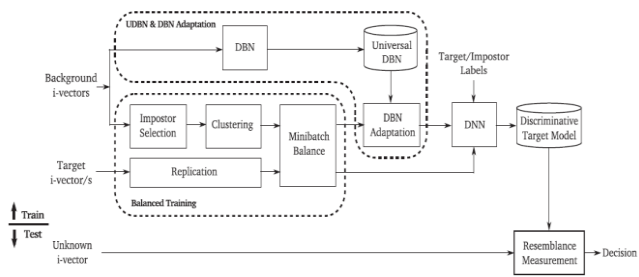

Fig. 7. Block-diagram of the train/test phases of the proposed deep learning backend for i-vectors in (Ghahabi and Hernando, 2017).

## 5.2.4. Using Contrastive loss for vector comparison

Since speaker identification is treated as a simple classification task, softmax layers can be applied to create a DNN backend system. However, in speaker verification, the comparison of two (speaker modelling) vectors is necessary. In a DNN, a way by this can be achieved is using contrastive loss (Chopra et al., 2005) as loss function on deep features. Convolutional networks (namely VGG (Simonyan and Zisserman, 2014; Yadav and Rai, 2018)) (Nagrani et al., 2017) and ResNets (He et al., 2015; Chung et al., 2018) can be trained by this way to perform speaker verification tasks. On VoxCeleb and VoxCeleb2 datasets, lower EERs were obtained than in the case of standard i-vector PLDA systems: 8.8%, 7.8% and 3.95% EERs for i-vector, CNN and ResNet, respectively. However, in (Chung et al., 2018) ResNet and the baseline system were not trained on the same dataset (RestNet: VoxCeleb2, i-vector: VexCeleb1), therefore this increase could come from the effect of the larger audio material.

## 5.2.5. SincNet

Convolutional neural networks (CNNs) are also used in speaker recognition, using spectrograms (Nagrani et al., 2017; Ji et al., 2018; Hajavi and Etemad, 2019) or raw speech waveform as input (Ravanelli and Bengio, 2018; Salvati et al., 2019). SincNet (Ravanelli and Bengio, 2018) is a special CNN architecture that gets raw waveforms as inputs. Before applying standard CNN/DNN layers it learns high and low cut-off frequencies of band-pass filters by a convolutional layer (Fig. 8). In speaker identification task, compared to MFCC-fed DNN, the SincNet achieved better performance on TIMIT and LibriSpeech: 0.99%, 2.02% Classification Error Rate (CER) for TIMIT and LibriSpeech with DNN, and 0.85% and 0.96% CER for SincNet, respectively. SincNet was also compared to CNN with filterbank energies as inputs. The conclusion was that on smaller dataset (such as TIMIT), the filter learning was not as effective, as on a large dataset (LibriSpeech). On TIMIT, the results were comparable. On LibriSpeech however, SincNet outperformed the CNN architecture (1.55% and 0.96% CER for CNN and SincNet, respectively). It was found that SincNet al.,so outperformed the other DNN solutions (and the standard i-vector PLDA system) in a speaker verification setup. Both d-vector (used with cosine distance) and speaker class posteriors were applied.

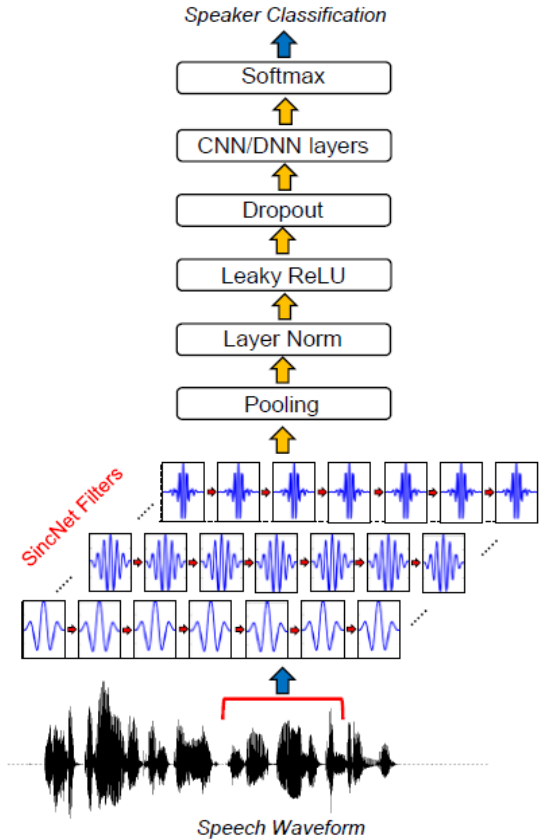

*Fig. 8. Architecture of SincNet in (Ravanelli and Bengio, 2018)*

SincNet was extended in (Ravanelli and Bengio, 2019) for an unsupervised speaker embedding learning by using mutual information as objective function for embedding vector comparison. An additional decrease in EER was examined: from 7.2 to 5.8% on the VoxCeleb corpus.

*5.2.6. Unlabeled data*

When doing speaker recognition, labeled data is not always present. There are some approaches that takes advantage of large scale unlabelled training data. Curriculum learning is one of them (Marchi et al., 2018; Ranjan and Hansen, 2018; Zheng et al., 2019). It starts by learning certain DNN model using a labeled corpus and continuously introduces unlabeled, out-of-domain text independent speaker samples. Both LSTM (Marchi et al., 2018) and TDNN (Zheng et al., 2019) based systems are proposed that outperform baseline methods.

*5.3. Other usage of DNN in speaker recognition*

In (Lei et al., 2014) DNN is used in a non-common way to aid speaker recognition. The extraction of sufficient statistics for the general i-vector model is driven by a deep neural network trained for automatic speech recognition. This DNN is used to produce frame alignments, specifically providing posteriors of semitones. First, DNN is trained for segmenting the speech into senones, using a pre-trained general HMM-GMM ASR system. The i-vector training is done on the semitone-level segmented speech. The final flow diagram of the proposed method is shown in Fig. 6. The experiments were done using the two extended NIST SRE'12 conditions: clean and slightly noisy telephone speech. The pre-trained HMM-GMM system used a 39 dimensional MFCC vector, including 13 MFCC and their first and second derivatives. The input of the DNN in the HMM-DNN was composed of 15 frames, using 40 log Mel-filterbank for each. The results of the proposed method was compared to a standard i-vector system (GMM-UBM and i-vector). The HMM-DNN method achieved a slightly lower EER: 1.39% and 1.81% for DNN and UBM, respectively for clean speech; 1.92% and 2.55% for DNN and UBM, respectively for noisy speech.

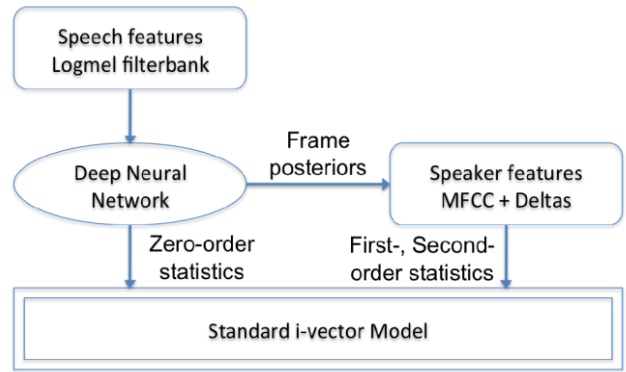

*Fig. 6. Flow diagram of the DNN/i-vector hybrid framework in (Lei et al., 2014)*

## 6 Conclusion

In this paper we tried to summarize the applied deep learning practices in the field of speaker recognition, both verification and identification. The early DL solutions to replace feature extraction (such as i-vectors) provided comparable but not higher performance than the previous state-of-the-art i-vector PLDA systems. Although, newer DL architectures lead to increasing classification accuracies, it is well-known in the literature that i-vectors provide competitive performance, when more training material is used for each speaker and when longer test sentences are employed (Sarkar et. al, 2012; Travadi et. al, 2014; Kanagasundaram et. al, 2011). However, the latest works offer superior results. In some cases, the reported results show significantly lower EERs, but mostly the achieved performances are only a little better than the previous ones. Nonetheless, it seems that DL becomes the now state-of-the-art solution for both speaker verification and identification. The standard x-vectors, additional to i-vectors, are used as baseline in most of the novel works. The increasing amount of gathered data opens up the territory to DL, where they are the most effective. Additionally, newer and newer DL architectures are developed, that can lead to a breakthrough in speaker recognition too.

## Acknowledgements

The work was funded by project no. FK128615, which has been implemented with the support provided from the National Research, Development and Innovation Fund of Hungary, financed under the FK_18 funding scheme.